\documentclass[nonacm,sigconf,bookmarksnumbered,unicode,bookmarks=false,hidelinks]{acmart}

\pdfoutput=1
\usepackage{etex}

\copyrightyear{2020} 
\acmYear{2020} 
\setcopyright{acmlicensed}
\acmConference[ICSE '20]{42nd International Conference on Software Engineering}{May 23--29, 2020}{Seoul, Republic of Korea}
\acmBooktitle{42nd International Conference on Software Engineering (ICSE '20), May 23--29, 2020, Seoul, Republic of Korea}
\acmPrice{15.00}
\acmDOI{10.1145/3377811.3380364}
\acmISBN{978-1-4503-7121-6/20/05}

\usepackage[utf8]{inputenc}
\usepackage[english]{babel}
\usepackage[T1]{fontenc}
\usepackage{url}
\usepackage[keeplastbox]{flushend}
\usepackage{xspace}
\usepackage{caption}
\usepackage[autolanguage]{numprint}
\usepackage[inline]{enumitem}
\usepackage{ifthen}
\usepackage{colortbl}
\usepackage{eurosym}
\usepackage{adjustbox}

\usepackage{graphicx}
\usepackage[caption=false]{subfig}
\usepackage{pgfplots}
\usepackage{pgf}
\pgfplotsset{compat=1.14} 
\usetikzlibrary{pgfplots.groupplots}

\usepackage{booktabs}
\usepackage{multirow}
\usepackage{threeparttable}
\usepackage{xcolor}
\usepackage{array,makecell}
\usepackage{diagbox}
\usepackage{tabularx}

\definecolor{bblue}{HTML}{4F81BD}
\definecolor{rred}{HTML}{C0504D}
\definecolor{ggreen}{HTML}{9BBB59}
\definecolor{ppurple}{HTML}{9F4C7C}

\usepackage{fp}
\newcommand{\ShowAbsoluteNumber}[1]{%
\ifnum #1<10%
{\hspace*{0pt}#1}%
\else%
\ifnum #1<100%
{\hspace*{0pt}#1}%
\else%
\ifnum #1<1000%
{\hspace*{0pt}#1}%
\else%
{\numprint{#1}}%
\fi%
\fi%
\fi%
}

\newcommand{\ShowPercentage}[2]{%
\FPeval\percentage{round(#1/#2*100,0)}%
\ifnum \percentage=0%
{\scriptsize(<1\%)}%
\else%
\ifnum \percentage<10%
{\small(\hspace*{0pt}\FPprint{percentage}\%)}%
\else%
{\small(\FPprint{percentage}\%)}%
\fi%
\fi%
\hspace*{0.3ex}%
}
\newlength\BARSIZE  
\newcommand{\inlinechart}[2]{%
\FPeval{\BLACKBARSIZE}{#1/#2}\textcolor{black!80}{\rule{\BLACKBARSIZE\BARSIZE}{1.6ex}}%
\FPeval{\BLACKBARSIZE}{1 - (#1/#2)}\textcolor{black!10}{\rule{\BLACKBARSIZE\BARSIZE}{1.6ex}}%
}

\newcommand*\ChartSmall[2]{%
$#1$\hspace*{0.5ex}%\ShowPercentage{#1}{#2}%
\inlinechart{#1}{#2}%
}

\newcommand{\answer}[2]{\vspace{.1cm}{\centering\fbox{\parbox{0.97\columnwidth}{\textbf{Answer to RQ#1}. #2}}}\vspace{.1cm}}

\newboolean{showcomments}
\setboolean{showcomments}{true}

\ifthenelse{\boolean{showcomments}}
  {\newcommand{\nb}[3]{
  {\color{#2}\small\fbox{\bfseries\sffamily\scriptsize#1}}
  {\color{#2}\sffamily\small$\triangleright~$\textit{\small #3}$~\triangleleft$\GenericWarning{}{LaTeX Warning: #1: #3}}
  }
  \newcommand{\todo}[1]{{\color{red}{TODO: #1}}\GenericWarning{}{LaTeX Warning: TODO: #1}}
  }
  {\newcommand{\nb}[3]{}
  \newcommand{\todo}[1]{}
  }
  
\NewEnviron{autoscaletikz}%
{\noindent\hphantom{\vphantom{\begin{tikzpicture}\BODY\getxyscale\end{tikzpicture}}}%
\noindent\begin{tikzpicture}[xscale=\xscalefactor,yscale=\yscalefactor]\BODY\end{tikzpicture}}

\def\nbTotalTools{35\xspace}%
\def\nbTools{9\xspace}%
\def\nbAnalisis{{428,337}\xspace}%
\def\nbCurated{{69}\xspace}%
\def\nbContracts{{47,518}\xspace}%
\def\nbContractsAll{{47,587}\xspace}%
\def\nbDuplicatedContracts{{972,975}\xspace}%
\def\ExecDuration{564 days and 3 hours\xspace}

\def\smartbugs{SmartBugs\xspace}

\newcommand\sbset[1]{$\textsc{sb}^{\textsc{#1}}$\xspace}
\def\sbcurated{\sbset{curated}}
\def\sbwild{\sbset{wild}}

\begin{document}

\title{Empirical Review of Automated Analysis Tools on \nbContractsAll Ethereum~Smart Contracts}

\author{Thomas Durieux}
\affiliation{%
    \institution{INESC-ID and IST, University of Lisbon, Portugal}
    %\country{Portugal}
}
\email{thomas@durieux.me}

\author{Jo\~ao F. Ferreira}
\affiliation{%
    \institution{INESC-ID and IST, University of Lisbon, Portugal}
    %\country{Portugal}
}
%\email{joaofernandoferreira@tecnico.ulisboa.pt}
\email{joao@joaoff.com}

\author{Rui Abreu}
\affiliation{%
    \institution{INESC-ID and IST, University of Lisbon, Portugal}
    %\country{Portugal}
}
\email{rui@computer.org}

\author{Pedro Cruz}
\affiliation{%
    \institution{INESC-ID and IST, University of Lisbon, Portugal}
    %\institution{IST, University of Lisbon, Portugal}
    %\country{Portugal}
}
\email{pedrocrvz@gmail.com}

\begin{abstract}
Over the last few years, there has been substantial research on automated analysis, testing, and debugging of Ethereum smart contracts.
However, it is not trivial to compare and reproduce that research.
To address this, we present an empirical evaluation of \nbTools state-of-the-art automated analysis tools using two
new datasets: i) a dataset of 69 annotated vulnerable smart contracts that can be used to evaluate the precision
of analysis tools; and ii) a dataset with all the smart contracts in the Ethereum Blockchain that have 
Solidity source code available on Etherscan (a total of 47,518 contracts).
The datasets are part of \smartbugs, a new extendable execution framework that we created to facilitate the
integration and comparison between multiple analysis tools and the analysis of Ethereum smart contracts.
We used \smartbugs to execute the \nbTools automated analysis tools on the two datasets. In total,
we ran 428,337 analyses that took approximately 564 days and 3 hours, being
the largest experimental setup to date both in the number of tools and in execution time.
We found that only 42\% of the vulnerabilities from our annotated dataset are detected by all the tools, 
with the tool \textit{Mythril} having the higher accuracy (27\%).
When considering the largest dataset, we observed that 97\% of contracts are tagged as vulnerable, thus
suggesting a considerable number of false positives. Indeed, only a small number of vulnerabilities (and of only two categories) were detected simultaneously by four or more tools.
\end{abstract}

\begin{CCSXML}
<ccs2012>
<concept>
<concept_id>10011007.10011074.10011099.10011102</concept_id>
<concept_desc>Software and its engineering~Software defect analysis</concept_desc>
<concept_significance>500</concept_significance>
</concept>
<concept>
<concept_id>10011007.10011074.10011099.10011102.10011103</concept_id>
<concept_desc>Software and its engineering~Software testing and debugging</concept_desc>
<concept_significance>500</concept_significance>
</concept>
</ccs2012>
\end{CCSXML}

\ccsdesc[500]{Software and its engineering~Software defect analysis}
\ccsdesc[500]{Software and its engineering~Software testing and debugging}

\keywords{Smart contracts%
, Solidity%
, Ethereum%
, Blockchain%
%, Software%
, Tools%
, Debugging%
, Testing%
, Reproducible Bugs%
}

\maketitle

\section{Introduction}

% context
Blockchain technology has been receiving considerable attention from industry and 
academia, for it promises to disrupt the digital online world by enabling
a democratic, open, and scalable digital economy based on decentralized distributed
consensus without the intervention of third-party trusted authorities.
Among the currently available blockchain-based platforms,
Ethereum~\cite{buterin2013ethereum}  is one of the most popular, mainly because it enables developers to write distributed applications (Dapps) based on smart contracts\,---\,programs that are executed across a decentralised network of  nodes. The main language used to develop Ethereum smart contracts
is Solidity\footnote{Interested readers on Solidity, refer to 
\url{https://solidity.readthedocs.io}.},
a high-level language that follows a 
JavaScript-like, object-oriented paradigm. 
Contracts written in Solidity are compiled to bytecode that can be executed on the 
Ethereum Virtual Machine (EVM).

Smart contracts are at the core of Ethereum's value. 
However, as noted by some researchers~\cite{bhargavan2016formal,luu2016making}, 
due to the idiosyncrasies of the EVM, writing secure smart contracts is far from trivial.
In a preliminary study performed on nearly one million Ethereum smart contracts, using one 
analysis framework for verifying correctness, 
\emph{34,200} of them were flagged as vulnerable~\cite{nikolic2018finding}. 
Also, Luu \textit{et~al.}~\cite{luu2016making} proposed the symbolic execution tool Oyente and
showed that of \emph{19,366} Ethereum smart contracts analyzed, \emph{8,833} (around 46\%) 
were flagged as vulnerable. 
Famous attacks, such as TheDAO exploit~\cite{analysisthedao} and the Parity 
wallet bug~\cite{paritywallet} illustrate this problem and have led to huge financial 
losses.

% problem
There has been some effort from the research community to develop automated
analysis tools that locate and eliminate vulnerabilities in smart contracts~\cite{luu2016making,tikhomirov2018smartcheck,tsankov2018securify,grishchenko2018semantic}. 
However, it is not easy to compare and reproduce that research:
even though several of the tools are publicly available, the datasets used are not.
If a developer of a new tool wants to compare the new tool with
existing work, the current approach is to contact the authors of alternative
tools and hope that they give access to their datasets (as done in, e.g.,~\cite{perez2019smart}).

% our solution
{The aim of this paper is twofold. First, to be able to execute and compare automated
analysis tools, hence setting the ground for fair comparisons, we provide two datasets 
of Solidity smart contracts. The first dataset contains \nbCurated manually annotated smart contracts
that can be used to evaluate the precision of analysis tools. The second dataset contains all available 
smart contracts in the Ethereum Blockchain that have Solidity source code available on
Etherscan (a total of 47,518 contracts) at the time of writing.
We have executed \nbTools{} state-of-the-art automated analysis tools on the two datasets and analyzed the results 
in order to provide a fair point of comparison for future smart contract analysis tools.
In total, the execution of all the tools required 564 days and 3 hours to complete 
428,337 analyses. 

Second, to simplify research on automated analysis techniques for smart 
contracts, we provide a novel, extendable, and easy-to-use execution framework, called 
\smartbugs{}, to execute these tools on the same execution environment. This framework currently contains \nbTools{} configured smart contract analysis tools.}

In summary, the contributions of the paper are:
\begin{enumerate*}
    \item A dataset of annotated vulnerable Solidity smart contracts;
    \item A dataset that contains all the available smart contracts from the Ethereum blockchain that have Solidity source code available in Etherscan;
    \item An execution framework that includes \nbTools pre-configured smart contract analysis tools; and
    \item An analysis of the execution of \nbTools tools on \nbContractsAll smart contracts.
\end{enumerate*}

Our study demonstrates that there are several open challenges that need to be addressed by future work to improve the quality of existing tools and techniques.
We report that the current state-of-the-art is not able to detect vulnerabilities from 
two categories of DASP10: \textit{Bad Randomness} and \textit{Short Addresses}. Also, the 
tools are only able to detect together 42\% of the vulnerabilities from our 
dataset of annotated vulnerable smart contracts (48 out of 115). The most accurate tool, \textit{Mythril}, is able to detect only 27\% of the vulnerabilities.
%%% false positives
When considering the largest dataset, 97\% of contracts are tagged as vulnerable, thus suggesting a considerable number of false positives. %Indeed, only a small number of vulnerabilities (and of only two categories) were detected simultaneously by four or more tools.
In conclusion, we show that state-of-the-art techniques are far from being perfect, still likely producing too many false positives. On the positive side, the best performing techniques do so at a marginal execution cost.

\begin{center}
    \smartbugs is available at
    
    %\url{https://anonymous.4open.science/r/96fac4dd-5321-4abd-8399-246aac702fc6/}
    %\url{https://github.com/smartbugs/smartbugs}
    \url{https://smartbugs.github.io}
\end{center}

%The remainder of this paper is organized as follows.
%\autoref{sec:smartbugs:study-design} presents the design of our study, the tools 
%that we use in the study, the research questions, the data collection, and our 
%analysis methodology.
%\autoref{sec:smartbugs:results} presents our results.
%\autoref{sec:smartbugs:discussion} contains a discussion on practical implications and 
%challenges as well as the threats to validity.
%Finally, \autoref{sec:smartbugs:related-works} presents the related work, and
%\autoref{sec:smartbugs:conclusions} presents our final remarks.
%
\section{Study Design}
\label{sec:smartbugs:study-design}
Blockchain technologies are getting more and more attention from the research community and also, more importantly, from industry. As more blockchain-based solutions emerge, there is a higher reliance on the quality of the smart contracts. The industry and the research community came up with automatic approaches that analyze smart contracts to identify vulnerabilities and bad practices.
%The goals of this paper are to report the current state of the art of the current available automatic smart contracts analyzers and to provide an extendable execution framework that can easily execute analysis on smart contracts.
The main goal of this paper is to report the current state of the art of currently available automated analysis tools for smart contracts. To facilitate reproducibility and comparison between tools, the study is performed using a new extendable execution framework that we call \smartbugs (see \autoref{sec:smartbugs:framework}).

In this section, we present the design of our study, including the research questions, the systematic selection of the tools and datasets of smart contracts, the execution framework, and the data collection and analysis methodology.

\subsection{Research Questions}

In this study, we aim to answer the following research questions:

\begin{itemize}[leftmargin=24pt]
\item[\textbf{RQ1}.] {[Effectiveness]} What is the effectiveness of current analysis tools in detecting vulnerabilities in Solidity smart contracts?

In this first research question, we are interested in determining how precise state-of-the-art analysis tools are in detecting vulnerabilities in known faulty smart contracts.

\item[\textbf{RQ2}.] {[Production]} How many vulnerabilities are present in the Ethereum blockchain?

In this research question, we investigate the vulnerabilities that are detected in contracts pulled from the Ethereum blockchain.
We consider the most popular vulnerabilities, the evolution of the vulnerabilities over time, and the consensus among different combinations of automated analysis tools. %combining the tools to create a consensus between them. 

\item[\textbf{RQ3}.] {[Performance]} How long do the tools require to analyze the smart contracts?

And finally, we compare the performance of the analysis tools. The goal is to identify which tool is the most efficient.

\end{itemize}

\subsection{Subject Tools}
\label{sec:smartbugs:tools}
\begin{table}[t]
    \caption{Tools identified as potential candidates for this study.}
    \label{tab:smartbugs:analyzers}
    \centering
    \small
    \begin{tabularx}{0.47\textwidth}{@{}r l X@{}}
\toprule
 \# & Tools  & Tool URLs \\\midrule
1 & contractLarva \cite{azzopardi2018monitoring} & \url{https://github.com/gordonpace/contractLarva} \\
2 & E-EVM \cite{norvill2018visual} & \url{https://github.com/pisocrob/E-EVM} \\
3 & Echidna & \url{https://github.com/crytic/echidna} \\
4 & Erays \cite{zhou2018erays} & \url{https://github.com/teamnsrg/erays} \\
5 & Ether \cite{liu2018s} & \texttt{N/A} \\
6 & Ethersplay & \url{https://github.com/crytic/ethersplay} \\
7 & EtherTrust \cite{grishchenko2018ethertrust} & \url{https://www.netidee.at/ethertrust} \\
8 & EthIR \cite{albert2018ethir} & \url{https://github.com/costa-group/EthIR} \\
9 & FSolidM \cite{mavridou2018tool} & \url{https://github.com/anmavrid/smart-contracts} \\
10 & Gasper \cite{chen2017under} & \texttt{N/A} \\
11 & HoneyBadger \cite{honeybadger} & \url{https://github.com/christoftorres/HoneyBadger} \\
12 & KEVM \cite{hildenbrandt2018kevm} & \url{https://github.com/kframework/evm-semantics} \\
13 & MadMax \cite{grech2018madmax} & \url{https://github.com/nevillegrech/MadMax} \\
14 & Maian \cite{nikolic2018finding} & \url{https://github.com/MAIAN-tool/MAIAN} \\
15 & Manticore \cite{mossberg2019manticore}& \url{https://github.com/trailofbits/manticore/} \\
16 & Mythril \cite{mueller2018smashing} & \url{https://github.com/ConsenSys/mythril-classic} \\
17 & Octopus & \url{https://github.com/quoscient/octopus} \\
18 & Osiris \cite{torres2018osiris} & \url{https://github.com/christoftorres/Osiris} \\
19 & Oyente \cite{luu2016making} & \url{https://github.com/melonproject/oyente} \\
20 & Porosity \cite{suiche2017porosity} & \url{https://github.com/comaeio/porosity} \\
21 & rattle & \url{https://github.com/crytic/rattle} \\
22 & ReGuard \cite{liu2018reguard} & \texttt{N/A} \\
23 & Remix & \url{https://github.com/ethereum/remix} \\
24 & SASC \cite{zhou2018security} & \texttt{N/A} \\
25 & sCompile \cite{chang2018scompile} & \texttt{N/A} \\
26 & Securify \cite{tsankov2018securify} & \url{https://github.com/eth-sri/securify} \\
27 & Slither \cite{feistslither} & \url{https://github.com/crytic/slither} \\
28 & Smartcheck \cite{tikhomirov2018smartcheck} & \url{https://github.com/smartdec/smartcheck} \\
29 & Solgraph & \url{https://github.com/raineorshine/solgraph} \\
30 & Solhint & \url{https://github.com/protofire/solhint} \\
31 & SolMet \cite{hegedus2019towards} & \url{https://github.com/chicxurug/SolMet-Solidity-parser} \\
32 & teEther \cite{krupp2018teether} &  \url{https://github.com/nescio007/teether} \\
33 & Vandal \cite{brent2018vandal} & \url{https://github.com/usyd-blockchain/vandal} \\
34 & VeriSol \cite{microsoft_verisol} & \url{https://github.com/microsoft/verisol} \\
35 & Zeus \cite{kalra2018zeus} & \texttt{N/A} \\
 \bottomrule
    \end{tabularx}
\end{table}

\begin{table}[t]
    \small
    \caption{Excluded and included analysis tools based on our inclusion criteria.}
    \label{tab:smartbugs:selection-tools}
    \centering
    \begin{tabularx}{0.47\textwidth}{@{}p{.03\textwidth} l X@{}}
        \toprule
        {} & Inclusion criteria & Tools that violate criteria \\
        \midrule
        \multirow{8}{*}{\rotatebox[origin=c]{90}{\thead{Excluded\\(26)}}} &
             Available and CLI (C1) &  Ether, Gasper, ReGuard, Remix,  SASC, sCompile, teEther, Zeus\\
        {} & Compatible Input (C2) &  MadMax, Vandal\\
        {} & Only Source (C3) &  Echidna, VeriSol \\
        {} & Vulnerability Finding (C4) & contractLarva, E-EVM, Erays, Ethersplay, EtherTrust, EthIR, FSolidM, KEVM, Octopus, Porosity, rattle, Solgraph, SolMet, Solhint\\
        \midrule
        \rotatebox[origin=c]{90}{\thead{Included\\(\nbTools)}} & \multicolumn{2}{c}{\thead{HoneyBadger, Maian, Manticore, Mythril, Osiris, Oyente, \\Securify, Slither, Smartcheck}} \\
        \bottomrule
    \end{tabularx}
\end{table}

In order to discover smart contract automated analysis tools, we started off by using the survey of Angelo \textit{et~al.}~\cite{di2019survey} and we extended their list of tools by searching the academic literature and the internet for other tools. 
We ended up with the \nbTotalTools tools that are listed in \autoref{tab:smartbugs:analyzers}.

Not all the identified tools are well suited for our study. Only the tools that met the following three inclusion criteria were included in our study:

\begin{itemize}[leftmargin=*]
\item \textit{Criterion \#1.} [Available and CLI] The tool is publicly available and supports a command-line interface (CLI). The CLI facilitates the scalability of the analyses.

\item \textit{Criterion \#2.} [Compatible Input] The tool takes as input a Solidity contract. This excludes tools that only consider EVM bytecode.

\item \textit{Criterion \#3.} [Only Source] The tool requires only the source code of the contract to be able to run the analysis. This excludes tools that require a test suite or contracts annotated with assertions.

\item \textit{Criterion \#4.} [Vulnerability Finding] The tool identifies vulnerabilities or bad practices in contracts. This excludes tools that are described as analysis tools, but only construct artifacts such as control flow graphs.
\end{itemize}

After inspecting all \nbTotalTools analysis tools presented in \autoref{tab:smartbugs:analyzers}, we found \nbTools tools that meet the inclusion criteria outlined.
\autoref{tab:smartbugs:selection-tools} presents the excluded and included tools, and for the excluded ones, it also shows which criteria they did not meet. 
%\Joao{Placing Remix IDE as violating C4 is probably not accurate: Remix has plugins that do static analysis and identify problems / bad practices if I am not mistaken.}\Rui{Where would you place it?}

\textbf{HoneyBadger \cite{honeybadger}} is developed by a group of researchers at the University of Luxembourg and is an Oyente-based (see below) tool that employs symbolic execution and a set of heuristics to pinpoint honeypots in smart contracts. Honeypots are smart contracts that \emph{appear} to have an obvious flaw in their design, which allows an arbitrary user to drain Ether\footnote{Ether is the cryptocurrency of Ethereum.} from the contract, given that the user transfers a priori a certain amount of Ether to the contract. When HoneyBadger detects that a contract \emph{appears} to be vulnerable, it means that the developer of the contract wanted to make the contract look vulnerable, but is not vulnerable.

\textbf{Maian \cite{nikolic2018finding}}, developed jointly by researchers from the National University of Singapore and University College London, is also based on the Oyente tool. Maian looks for contracts that can be self-destructed or drained of Ether from arbitrary addresses, or that accept Ether but do not have a payout functionality. A dynamic analysis in a private blockchain is then used to reduce the number of false positives.

\textbf{Manticore \cite{mossberg2019manticore}}, developed by TrailOfBits, also uses symbolic execution to find execution paths in EVM  bytecode that lead to reentrancy vulnerabilities and reachable self-destruct operations.

\textbf{Mythril \cite{mueller2018smashing}}, developed by ConsenSys, relies on concolic analysis, taint analysis and control flow checking of the EVM bytecode to prune the search space and to look for values that allow exploiting vulnerabilities in the smart contract. 

\textbf{Osiris \cite{torres2018osiris}}, developed by a group of researchers at the University of Luxembourg, extends Oyente to detect integer bugs in smart contracts.

\textbf{Oyente \cite{luu2016making}}, developed by Melonport AG,  is one of the first smart contract analysis tools. It is also used as a basis for several other approaches like Maian and Osiris. Oyente 
%executes EVM bytecode using 
uses symbolic execution on EVM bytecode to identify vulnerabilities.
%analyze the trace and identify the vulnerabilities in the smart contracts.

\textbf{Securify \cite{tsankov2018securify}}, developed by ICE Center at ETH Zurich, statically analyzes EVM bytecode to infer relevant and precise semantic information about the contract using the Souffle Datalog solver. It then checks compliance and violation patterns that capture sufficient conditions for proving if a property holds or not.

\textbf{Slither \cite{feistslither}}, developed by TrailOfBits, is a static analysis framework that converts Solidity smart contracts into an intermediate representation called SlithIR and applies known program analysis techniques such as dataflow and taint tracking to extract and refine information.

\textbf{Smartcheck \cite{tikhomirov2018smartcheck}}, developed by SmartDec, is a static analysis tool that looks for vulnerability patterns and bad coding practices. It runs lexical and syntactical analysis on Solidity source code.

\subsection{Datasets of Smart Contracts}\label{sec:smartbugs:datasets}

For this study, we crafted two datasets of Solidity smart contracts with distinct purposes.
The first dataset, \sbcurated, consists of \nbCurated vulnerable smart contracts (see \autoref{sec:smartbugs:vulnerable_contracts}).
Contracts in this dataset are either real contracts that have been identified as vulnerable or have been purposely created to illustrate a vulnerability. 
%Those \nbCurated smart contracts have been identified as vulnerable or have been created to illustrate a vulnerability by a third party, such as by researchers or by the industry.
The goal of this dataset is to have a set of known vulnerable contracts labelled with the location and category of the vulnerabilities. This dataset can be used to evaluate the effectiveness of smart contract analysis tools in identifying vulnerabilities.

The second dataset is named \sbwild (see \autoref{sec:smartbugs:real_contracts}) and contains \nbContracts contracts extracted from the Ethereum blockchain. The set of vulnerabilities of those contracts is unknown; however, this dataset can be used to identify real contracts that have (potential) vulnerabilities and have an indication of how frequent a specific problem is. It can also be used to compare analysis tools in terms of metrics such as performance.

\subsubsection{\sbcurated: A Dataset of \nbCurated Vulnerable Smart Contracts}\label{sec:smartbugs:vulnerable_contracts}

\paragraph{Goal}

Our objective in constructing this dataset is to collect a set of Solidity smart contracts with known vulnerabilities, from deployed contracts in the Ethereum network to examples provided to illustrate vulnerabilities, that can serve as a dataset suite for research in the security analysis of Solidity smart contracts. 
We use the taxonomy presented in the DASP\footnote{Decentralized Application Security Project (or DASP): \url{https://dasp.co}} to describe vulnerabilities of Ethereum smart contracts (see Categories in \autoref{tab:selected-categories}). 
Each collected contract is classified in one of the ten categories. We also manually tagged the lines that contain the vulnerability. 
This classification allows for new smart contract analysis tools to be easily evaluated.

\paragraph{Collection Methodology}

\begin{table*}
\small
    \caption{Categories of vulnerabilities available in the dataset \sbcurated. For each category, we provide a description, the level at which the attack can be mitigated, the number of contracts available within that category, and the total number of lines of code in the contracts of that category (computed using cloc 1.82).}
    \label{tab:selected-categories}
    \centering
    \begin{tabularx}{\textwidth}{@{}p{9em}Xlrrr@{}}\toprule
            \textbf{Category} & \textbf{Description} & \textbf{Level} &
            \textbf{Contracts} & \textbf{Vulns} & \textbf{LoC}  \\
         \midrule
Access Control & Failure to use function modifiers or use of tx.origin & Solidity & 17 & 19 & \numprint{899}  \\
 
Arithmetic & Integer over/underflows & Solidity & 14 & 22 &\numprint{295}  \\

Bad Randomness &  Malicious miner biases the outcome & Blockchain & 8 & 31 &\numprint{1079} \\

Denial of service & The contract is overwhelmed with time-consuming computations & Solidity & 6 & 7 & 177  \\
 
Front running & Two dependent transactions that invoke the same contract are included in one block & Blockchain & 4 & 7 & 137 \\
 
Reentrancy & Reentrant function calls make a contract to behave in an unexpected way & Solidity & 7 & 8 & \numprint{778} \\

Short addresses & EVM itself accepts incorrectly padded arguments & EVM & 1 & 1 & 18 \\

Time manipulation & The timestamp of the block is manipulated by the miner & Blockchain & 4 & 5 & 76 \\

Unchecked low level calls & call(), callcode(), delegatecall() or send() fails and it is not checked & Solidity & 5 & 12  & 225 \\

Unknown Unknowns  & Vulnerabilities not identified in DASP 10 & \texttt{N/A} & 3  & 3 & 115 \\
         \midrule
         \textbf{Total} & & & \nbCurated & 115 & \numprint{3799} \\
    \bottomrule
    \end{tabularx}
\end{table*}

This dataset has been created by collecting contracts from three different sources: 1. GitHub repositories, 2. blog posts that analyze contracts, and 3. the Ethereum network.
80\% of the contracts were collected from GitHub repositories. 
We identified GitHub repositories that match relevant search queries 
(\emph{`vulnerable smart contracts'}, \emph{`smart contracts security'}, 
\emph{`solidity vulnerabilities'}) and contain 
vulnerable smart contracts. We searched Google using the same queries.
We found several repositories with vulnerable smart contracts such as \emph{not-so-smart-contracts}\footnote{not-so-smart-contracts: \url{https://github.com/crytic/not-so-smart-contracts}} and \emph{SWC Registry}\footnote{SWC Registry: \url{https://smartcontractsecurity.github.io/SWC-registry}}. The latter is a classification scheme for security weaknesses in Ethereum smart contracts that is referenced in Mythril's GitHub repository.
We also extracted vulnerabilities that come from trusted entities in blog posts where smart contracts are audited, tested or discussed such as Positive.com\footnote{Positive.com: \url{https://blog.positive.com}} and Blockchain.unica\footnote{Blockchain.unica: \url{http://blockchain.unica.it/projects/ethereum-survey/}}.
And finally, we used Etherscan\footnote{Etherscan: \url{https://etherscan.io}} to collect smart contracts that are deployed on the Ethereum network and are known to contain vulnerabilities (e.g. the original SmartBillions contract). 
Note that all the contracts were collected from trusted entities in the 
field. We also ensure the traceability of each contract by providing the URL from which they were taken and its author, where possible.

\paragraph{Dataset Statistics}

The dataset contains \nbCurated contracts and 115 tagged vulnerabilities, divided into ten categories of vulnerabilities.
\autoref{tab:selected-categories} presents information about the \nbCurated contracts. 
Each line contains a category of vulnerability.
For each category, we provide a description, the level at which the attack can be mitigated, the number of contracts available within that category, and the total number of lines of code in the contracts of that category.

\paragraph{Dataset Availability}
The dataset is available in the repository of \smartbugs \cite{Repo}. 
%The dataset is structured as follow. 
The dataset is divided into ten folders named with the DASP categories, and the folders contain the contracts of that category.
Moreover, the dataset contains the file \texttt{vulnerabilities.json} which contains the details of each vulnerable contract. It details the name, the origin URL, the path, and the lines and the category of the vulnerabilities.

\subsubsection{\sbwild: \nbContracts Contracts from the Ethereum Blockchain}\label{sec:smartbugs:real_contracts}

\paragraph{Goal}
The goal of this second dataset is to collect as many smart contracts as possible from the Ethereum blockchain, in order to have a representative picture of the practice and (potential) vulnerabilities that are present in the production environment.

\paragraph{Collection Methodology}
The data collection for the second dataset follows a different strategy. 
In this dataset, we collect all the different contracts from the Ethereum blockchain. Etherscan allows downloading the source code of a contract if you know its address.
Therefore, we firstly use Google BigQuery \cite{bigquery} to collect the Ethereum contract addresses that have at least one transaction. We used the following BigQuery request to select all the contract addresses and count the number of transactions that are associated with each contract\footnote{The query is also available at the following URL: \url{https://bigquery.cloud.google.com/savedquery/281902325312:47fd9afda3f8495184d98db6ae36a40c}}:
{\small
\begin{verbatim}
SELECT contracts.address, COUNT(1) AS tx_count
  FROM `ethereum_blockchain.contracts` AS contracts
  JOIN `ethereum_blockchain.transactions` AS transactions 
        ON (transactions.to_address = contracts.address)
  GROUP BY contracts.address
  ORDER BY tx_count DESC
\end{verbatim}
}

After collecting all the contract addresses, we used Etherscan and its API to 
retrieve the source code associated with an address.
However, Etherscan does not have the source code for every contract.
Therefore, at the end of this step, we obtained a Solidity file for each contract that 
has its source code available in the Etherscan platform.

The final step was to filter the set of contracts to remove duplicates. 
Indeed, we observe that 95\% of the available Solidity contracts are duplicates.
We consider that two contracts are duplicates when the MD5 checksums of the two source files are identical after removing all the spaces and tabulations.

\paragraph{Dataset Statistics}

\begin{table}[t]
    \small
    \caption{Statistics on the collection of Solidity smart contracts from the Ethereum blockchain.}
    \label{tab:smartbugs:collection_stat}
    \centering
    \begin{tabularx}{0.45\textwidth}{@{}X r@{}}\toprule
Solidity source not available & \numprint{1290074} \\
Solidity source available & \numprint{972975} \\
Unaccessible & 47  \\ \midrule
\textbf{Total} & \numprint{2263096} \\\midrule
\textbf{Unique Solidity Contracts} & \nbContracts \\
LOC & \numprint{9693457}\\
% Version & 359 \\
% Names & 30417 \\
\bottomrule
\end{tabularx}

\end{table}
\autoref{tab:smartbugs:collection_stat} presents the statistics of this dataset.
The query on Google BigQuery retrieved \numprint{2263096} smart contract addresses.
We then requested Etherscan for the Solidity source code of those contracts, and we obtained \numprint{972975} Solidity files. This means that
\numprint{1290074} of the contracts do not have an associated source file in Etherscan.
The filtering process of removing duplicate contracts resulted in \nbContracts unique contracts (a total of 9,693,457 lines). According to Etherscan, 47 contracts that we requested do not exist (we labelled those as Unaccessible).

\paragraph{Dataset Availability}

The dataset is available on GitHub \cite{WildDataset}. The dataset contains the Solidity source code of each of the \nbContracts contracts.
The contracts are named with the address of the contract.
We also attached with this dataset additional information in order to use this dataset for other types of studies. 
It contains:
\begin{itemize*}
    \item the name of the contract;
    \item the Solidity version that has been used to compile the contract;
    \item the addresses of the duplicated contracts;
    \item the number of transactions associated with the contract; %that the contracts did;
    \item the size of the contract in lines of Solidity code;
    \item the date of the last transactions for the \numprint{2263096} contracts;
    \item the date of creation for the \numprint{2263096} contracts; and
    \item the Ethereum balance of \nbDuplicatedContracts contracts that have their source code available.
\end{itemize*}

\subsection{The Execution Framework: \smartbugs}\label{sec:smartbugs:framework}

We developed \smartbugs, an execution framework aiming at simplifying the execution of
analysis tools on datasets of smart contracts.
\smartbugs has the following features:
\begin{itemize*}
    \item A plugin system to easily add new analysis tools, based on Docker images;
    \item Parallel execution of the tools to speed up the execution time;
    \item An output mechanism that normalizes the way the tools are outputting the results, and simplify the process of the output across tools.
\end{itemize*}

\smartbugs currently supports \nbTools tools (see \autoref{sec:smartbugs:tools}).

\subsubsection{Architecture}

% \autoref{fig:smartbugs:smartbugs_architecture} presents the architecture of \smartbugs. 
\smartbugs is composed of five main parts:
\begin{enumerate*}
\item The first consists of the command-line interface to use \smartbugs (see \autoref{sec:smartbugs:interface}).
\item The second part contains the tool plugins. Each tool plugin contains the configuration of the tools. The configuration contains the name of the Docker image, the name of the tool, the command line to run the tool, the description of the tool, and the location of the output of results.
\item The Docker images that are stored on Docker Hub. We use Docker images of the tools when a Docker image is already available; otherwise, we create our own image (all Docker images are publicly available on Docker Hub, including our own).
\item The datasets of smart contracts (see \autoref{sec:smartbugs:datasets}).
\item The \smartbugs' runner puts all the parts of \smartbugs together to execute the analysis tools on the smart contracts.
\end{enumerate*}

% \begin{figure}
%     \centering
%     \includegraphics[width=.44\textwidth]{smartbugs_architecture.pdf}
%     \caption{Architecture of \smartbugs}
%     \label{fig:smartbugs:smartbugs_architecture}
% \end{figure}

\subsubsection{Dataset Interface Details}
\label{sec:smartbugs:interface}
\smartbugs provides a command-line interface that simplifies the execution of the smart contract analysis tools. It takes a set of tool names and a path to Solidity files to analyze and produces two files per execution: 1) a \texttt{result.log} file that contains the stdout of the execution and 2) a \texttt{result.json} file that contains the results of the analysis in a parsable format. Moreover, we provide scripts that process those outputs and render them in readable tables such as the one presented in this paper.

\subsection{Data Collection and Analysis}
To answer our research questions, we used \smartbugs to execute the \nbTools tools on the two datasets described in \autoref{sec:smartbugs:datasets}. We collected the output and used it for further analysis.
%resulting in vulnerabilities analyzes that are further used for analysis \Rui{strange sentence}.
In this section, we describe the setup of the tools (\autoref{sec:smartbugs:setup}) and their execution (\autoref{sec:smartbugs:execution}).

\subsubsection{Tools' Setup}\label{sec:smartbugs:setup}
For this experiment, we set the time budget to 30 minutes per analysis.
In order to identify a suitable time budget for one execution of one tool over one contract, we first executed all the tools on \sbcurated{} dataset. We then selected a time budget that is higher than the average execution time (one minute and 44 seconds). 
If the time budget is spent, we stop the execution and collect the partial results of the execution.
During the execution of our experiment, Manticore was the only tool that faced timeouts.
%no other tool but Manticore faced timeouts.

\subsubsection{Large-scale Execution}\label{sec:smartbugs:execution}

To our knowledge, we present the largest experimental study 
%our experimental setup is the largest one 
on smart contract analysis, both in the number of tools and in execution time.
In total, we executed \nbTools analysis tools on \nbContracts contracts.
This represents \nbAnalisis analyzes, which took approximately \ExecDuration of combined execution, more than a year of continuous execution.
We used two cloud providers to rent the servers required for this experiment. The first provider was Scaleway\footnote{Scaleway: \url{https://www.scaleway.com}}, where we used three servers with 32 vCPUs with 128 GB of RAM. We added a budget of \EUR{500}, and we spent \EUR{474.99}.

The second provider was Google Cloud\footnote{Google Cloud: \url{https://cloud.google.com}}, where we also used three servers with 32 vCPUs with 30GB of RAM. We spent \EUR{1038.46} with Google Cloud.
In total, we spent %\numprint{1038.46} + \numprint{474.99} = 
\EUR{1513.45} to execute the experiments discussed in this paper.
We used two cloud providers due to administrative restrictions on our budget line. We were initially targeting Scaleway because it is cheaper than Google Cloud, but we were not able to spend more than \EUR{500} with this provider.
All the logs and the raw results of the analysis are available at \cite{Results}.

\section{Results}\label{sec:smartbugs:results}

The results of our empirical study, as well as the answers to our research questions, are presented in this section.

\subsection{Precision of the Analysis Tools (RQ1)} \label{sec:smartbugs:rq1}

\begin{table*}
    \setlength{\tabcolsep}{3pt}
    \caption{Vulnerabilities identified per category by each tool. The number of vulnerabilities identified by a single tool is shown in brackets.}
    \label{tab:smartbugs:flagged-categories}
    \small
    
    \centering
    \begin{tabularx}{\textwidth}{@{}Xrrrrrrrrrr@{}}\toprule
Category           & HoneyBadger   & Maian         & Manticore     & Mythril       & Osiris        & Oyente        & Securify      & Slither       & Smartcheck    & Total         \\\midrule
% Access Control      &   0/19   0\% &   0/19   0\% &   4/19  21\% &   4/19  21\% &   0/19   0\% &   0/19   0\% &   0/19   0\% &   4/19  21\% &   2/19  11\% &   5/19  26\% \\
% Arithmetic          &   0/22   0\% &   0/22   0\% &   4/22  18\% &  15/22  68\% &  11/22  50\% &  12/22  55\% &   0/22   0\% &   0/22   0\% &   1/22   5\% &  19/22  86\% \\
% Denial of Service      &    0/7   0\% &    0/7   0\% &    0/7   0\% &    0/7   0\% &    0/7   0\% &    0/7   0\% &    0/7   0\% &    0/7   0\% &    0/7   0\% &   0/ 7   0\% \\
% Front Running       &    0/7   0\% &    0/7   0\% &    0/7   0\% &    2/7  29\% &    0/7   0\% &    0/7   0\% &    2/7  29\% &    0/7   0\% &    0/7   0\% &   2/ 7  29\% \\
% Reentrancy          &    0/8   0\% &    0/8   0\% &    2/8  25\% &    5/8  62\% &    5/8  62\% &    5/8  62\% &    5/8  62\% &    7/8  88\% &    5/8  62\% &   7/ 8  88\% \\
% Time Manipulation   &    0/5   0\% &    0/5   0\% &    1/5  20\% &    0/5   0\% &    0/5   0\% &    0/5   0\% &    0/5   0\% &    2/5  40\% &    1/5  20\% &   3/ 5  60\% \\
% Unchecked Low~Level~Calls &   0/12   0\% &   0/12   0\% &   2/12  17\% &   5/12  42\% &   0/12   0\% &   0/12   0\% &   3/12  25\% &   4/12  33\% &   4/12  33\% &   9/12  75\% \\
% Unknown~Unknowns               &    2/3  67\% &    0/3   0\% &    0/3   0\% &    0/3   0\% &    0/3   0\% &    0/3   0\% &    0/3   0\% &    3/3 100\% &    0/3   0\% &   3/ 3 100\% \\\midrule
Access Control      &   0/19   0\%     &   0/19   0\%     &   4/19  21\%     &   4/19  21\%     &   0/19   0\%     &   0/19   0\%     &   0/19   0\%     &   4/19  21\% (1) &   2/19  11\%     &   5/19  26\% \\
Arithmetic          &   0/22   0\%     &   0/22   0\%     &   4/22  18\%     &  15/22  68\%     &  11/22  50\% (2) &  12/22  55\% (2) &   0/22   0\%     &   0/22   0\%     &   1/22   5\%     &  19/22  86\% \\
Denial Service      &    0/7   0\%     &    0/7   0\%     &    0/7   0\%     &    0/7   0\%     &    0/7   0\%     &    0/7   0\%     &    0/7   0\%     &    0/7   0\%     &    0/7   0\%     &   0/ 7   0\% \\
Front Running       &    0/7   0\%     &    0/7   0\%     &    0/7   0\%     &    2/7  29\%     &    0/7   0\%     &    0/7   0\%     &    2/7  29\%     &    0/7   0\%     &    0/7   0\%     &   2/ 7  29\% \\
Reentrancy          &    0/8   0\%     &    0/8   0\%     &    2/8  25\%     &    5/8  62\%     &    5/8  62\%     &    5/8  62\%     &    5/8  62\%     &    7/8  88\% (2) &    5/8  62\%     &   7/ 8  88\% \\
Time Manipulation   &    0/5   0\%     &    0/5   0\%     &    1/5  20\%     &    0/5   0\%     &    0/5   0\%     &    0/5   0\%     &    0/5   0\%     &    2/5  40\% (1) &    1/5  20\% (1) &   3/ 5  60\% \\
Unchecked Low Calls &   0/12   0\%     &   0/12   0\%     &   2/12  17\%     &   5/12  42\% (1) &   0/12   0\%     &   0/12   0\%     &   3/12  25\%     &   4/12  33\% (3) &   4/12  33\% (1) &   9/12  75\% \\
Other               &    2/3  67\%     &    0/3   0\%     &    0/3   0\%     &    0/3   0\%     &    0/3   0\%     &    0/3   0\%     &    0/3   0\%     &    3/3 100\% (1) &    0/3   0\%     &   3/ 3 100\% \\
\midrule
Total               &  2/115   2\% &  0/115   0\% & 13/115  11\% & 31/115  27\% & 16/115  14\% & 17/115  15\% & 10/115   9\% & 20/115  17\% & 13/115  11\% & 48/115  42\% \\
    \bottomrule
    \end{tabularx}
\end{table*}

\begin{table*}
\caption{Total number of detected vulnerabilities by each tool, including vulnerabilities 
%that have not been 
not tagged in the dataset.}\label{tab:smartbugs:curved-all-detection}
 \small
\centering
\begin{tabularx}{\textwidth}{@{}Xrrrrrrrrrr@{}}
\toprule
Category            & HoneyBadger &    Maian    &  Manticore  &   Mythril   &   Osiris    &   Oyente    &  Securify   &   Slither   & Smartcheck  &    Total    \\\midrule
Access Control      &\ChartSmall{0}{91} &\ChartSmall{10}{91} &\ChartSmall{28}{91} &\ChartSmall{24}{91} &\ChartSmall{0}{91} &\ChartSmall{0}{91} &\ChartSmall{6}{91} &\ChartSmall{20}{91} &\ChartSmall{3}{91} & \ChartSmall{91}{769} \\
Arithmetic          &\ChartSmall{0}{257} &\ChartSmall{0}{257} &\ChartSmall{11}{257} &\ChartSmall{92}{257} &\ChartSmall{62}{257} &\ChartSmall{69}{257} &\ChartSmall{0}{257} &\ChartSmall{0}{257} &\ChartSmall{23}{257} & \ChartSmall{257}{769} \\
Denial of Service      &\ChartSmall{0}{59} &\ChartSmall{0}{59} &\ChartSmall{0}{59} &\ChartSmall{0}{59} &\ChartSmall{27}{59} &\ChartSmall{11}{59} &\ChartSmall{0}{59} &\ChartSmall{2}{59} &\ChartSmall{19}{59} & \ChartSmall{59}{769} \\
Front Running       &\ChartSmall{0}{76} &\ChartSmall{0}{76} &\ChartSmall{0}{76} &\ChartSmall{21}{76} &\ChartSmall{0}{76} &\ChartSmall{0}{76} &\ChartSmall{55}{76} &\ChartSmall{0}{76} &\ChartSmall{0}{76} & \ChartSmall{76}{769} \\
Reentrancy          &\ChartSmall{0}{84} &\ChartSmall{0}{84} &\ChartSmall{4}{84} &\ChartSmall{16}{84} &\ChartSmall{5}{84} &\ChartSmall{5}{84} &\ChartSmall{32}{84} &\ChartSmall{15}{84} &\ChartSmall{7}{84} & \ChartSmall{84}{769} \\
Time Manipulation   &\ChartSmall{0}{20} &\ChartSmall{0}{20} &\ChartSmall{4}{20} &\ChartSmall{0}{20} &\ChartSmall{4}{20} &\ChartSmall{5}{20} &\ChartSmall{0}{20} &\ChartSmall{5}{20} &\ChartSmall{2}{20} & \ChartSmall{20}{769} \\
Unchecked Low~Level~Calls &\ChartSmall{0}{82} &\ChartSmall{0}{82} &\ChartSmall{4}{82} &\ChartSmall{30}{82} &\ChartSmall{0}{82} &\ChartSmall{0}{82} &\ChartSmall{21}{82} &\ChartSmall{13}{82} &\ChartSmall{14}{82} & \ChartSmall{82}{769} \\
Unknown Unknowns               &\ChartSmall{5}{100} &\ChartSmall{2}{100} &\ChartSmall{25}{100} &\ChartSmall{32}{100} &\ChartSmall{0}{100} &\ChartSmall{0}{100} &\ChartSmall{0}{100} &\ChartSmall{28}{100} &\ChartSmall{8}{100} & \ChartSmall{100}{769} \\\midrule
Total               & \ChartSmall{5}{769} & \ChartSmall{12}{769} & \ChartSmall{76}{769} & \ChartSmall{215}{769} & \ChartSmall{98}{769} & \ChartSmall{90}{769} & \ChartSmall{114}{769} & \ChartSmall{83}{769} & \ChartSmall{76}{769} &769 \\
\bottomrule
\end{tabularx}
\end{table*}

To answer the first research question, we used \sbcurated{}, the dataset of \nbCurated contracts described in \autoref{sec:smartbugs:vulnerable_contracts}.
Since each contract of this dataset is categorized in one of the ten DASP categories, we can compute the ability of the \nbTools tools in detecting the vulnerabilities present in the \nbCurated contracts.
The methodology that we followed to answer this research question was the following:
%% I itemized the following steps; if we need space we can move them inline
\begin{enumerate*}
\item We executed the \nbTools tools on the \nbCurated contracts. The result of this execution is available on GitHub \cite{Results}.
\item We extracted all the vulnerabilities that were detected by the tools into a JSON file.
\item We mapped the detected vulnerabilities to a category of vulnerabilities (see \autoref{tab:selected-categories}). To achieve this task, we manually annotated all the vulnerability types that have been detected into one of the ten DASP categories. For example, Oyente detects a vulnerability called \texttt{Integer Overflow} that we link to the category \textit{Arithmetic}. In total, we identify 141 vulnerability types, and 97 of them have been tagged in one of the ten categories. The remaining 44 do not fit the DASP taxonomy (for example, some tools warn about the use of inline assembly, which is considered a bad practice but does not necessarily lead to vulnerable contracts)\footnote{The mapping that we created is available at: \url{https://github.com/smartbugs/smartbugs/wiki/Vulnerabilities-mapping}.}
\item At this point, we were able to identify which vulnerabilities the tools detect.
Unfortunately, we found out that none of the \nbTools tools were able to detect vulnerabilities of the categories \textit{Bad Randomness} and \textit{Short Addresses}.
%\todo{(JFF, R3Q2) The tools were not designed to detect those vulns.}
This is unsurprising: it is expected that tools do not detect vulnerabilities of certain categories, since they are not designed to identify all types of vulnerabilities. Despite not seeing this as a limitation of the studied tools, we argue that our analysis gives 
insight on opportunities to improve them, since it provides an overview on how tools perform with respect to the taxonomy used.
\end{enumerate*}

The results of this first study are presented in \autoref{tab:smartbugs:flagged-categories} and \autoref{tab:smartbugs:curved-all-detection}.
The first table presents the number of known vulnerabilities that have been identified. A vulnerability is considered as identified when a tool detects a vulnerability of a specific category at a specific line, and it matches the vulnerability that has been annotated in the dataset.
Each row of \autoref{tab:smartbugs:flagged-categories} represents a vulnerability category, and each cell presents the number of vulnerabilities where the tool detects a vulnerability of this category.
Some cells in \autoref{tab:smartbugs:flagged-categories} have numbers enclosed in brackets: these denote the number of vulnerabilities identified by a single tool.
This table summarizes the strengths and weaknesses of the current state of the art of smart contract analysis tools.
It shows that the tools can accurately detect vulnerabilities of the categories \textit{Arithmetic}, \textit{Reentrancy}, \textit{Time manipulation}, \textit{Unchecked Low Level Calls}, and \textit{Unknown Unknowns}. With respect to the category \textit{Unknown Unknowns}, the tools detected vulnerabilities such as the presence of uninitialized data and the possibility of locking down Ether.
However, they were not accurate in detecting vulnerabilities of the categories \textit{Access Control}, \textit{Denial of service}, and \textit{Front running}. The categories \textit{Bad Randomness} and \textit{Short Addresses} are not listed, since none of the tools are able to detect vulnerabilities of these types. 
This shows that there is still room for improvement and, potentially, for new approaches to detect vulnerabilities of the ten DASP categories.

\autoref{tab:smartbugs:flagged-categories} also shows that the tools offer distinct accuracies. Indeed, the tool \textit{Mythril} has the best accuracy among the \nbTools tools.
\textit{Mythril} detects 27\% of all the vulnerabilities when the average of all tools is 12\%. 
The ranking of the tools is comparable to the one observed by Parizi \textit{et~al.}~\cite{Parizi:2018:EVA:3291291.3291303}. However, the average accuracy is lower on our benchmark \sbcurated{}.
Moreover, \textit{Mythril},  \textit{Manticore}, \textit{Slither}, and \textit{Smartcheck} are the tools that detect the largest number of different categories (5 categories). We can also see that \textit{Slither} is the tool that uniquely identifies more vulnerabilities (8 vulnerabilities across 5 categories).
Despite its good results, \textit{Mythril} is not powerful enough to replace all the tools: by combining the detection abilities of all the tools, we succeed to detect 42\% of all the vulnerabilities. However, depending on the available computing power, it might not be realistic to combine all the tools.

Therefore, we suggest the combination of \textit{Mythril} and \textit{Slither}.
This combination detects 42 (37\%) unique vulnerabilities.
This combination offers a good balance between performance and execution cost.
This combination is the best possible combination by a considerable margin. The second best combination, \textit{Mythrill} and \textit{Oyente}, only succeeds to detect 33 (29\%) of all the vulnerabilities.

We now consider all the vulnerability detections and not only the ones that have been tagged in \sbcurated{}.
\autoref{tab:smartbugs:curved-all-detection} presents the total number of vulnerabilities detected by the tools.
This table allows the comparison of the total number of detected vulnerabilities with the number of detected known vulnerabilities shown in \autoref{tab:smartbugs:flagged-categories}. 
%Unsurprisingly, the more accurate a tool is in detecting known vulnerabilities, the more accurate it is at detecting unknown vulnerabilities.
The tools that are performing the best are also producing much more warnings (i.e., their output is more \textit{noisy}), making it difficult for a developer to exploit their results.

\answer{1}{\textbf{What is the accuracy of current analysis tools in detecting vulnerabilities on Solidity smart contracts?}
By combining the \nbTools tools together, they are only able to detect 42\% of all the vulnerabilities. 
This shows that there is still room to improve the accuracy of the current approaches to detect vulnerabilities in smart contracts.
We observe that the tools underperform to detect vulnerabilities in the following three categories: \textit{Access Control}, \textit{Denial of service}, and \textit{Front running}. They are unable to detect by design vulnerabilities from \textit{Bad Randomness} and \textit{Short Addresses} categories.
We also observe that \textit{Mythril} outperforms the other tools by the number of detected vulnerabilities (31/115, 27\%) and by the number of vulnerability categories that it targets (5/9 categories).
The combination of \textit{Mythril} and \textit{Slither} allows detecting a total of 42/115 (37\%) vulnerabilities, which is the best trade-off between accuracy and execution costs.
}

\subsection{Vulnerabilities in Production Smart Contracts (RQ2)}

\begin{table*}[t]
\caption{The total number of contracts that have at least one vulnerability (analysis of \nbContracts contracts).}
\label{tab:smartbugs:wild-all-detection}
\small
\centering
\begin{tabularx}{\textwidth}{@{}Xrrrrrrrrrr@{}}
\toprule
Category            & HoneyBadger &    Maian    &  Manticore  &   Mythril   &   Osiris    &   Oyente    &  Securify   &   Slither   & Smartcheck  &    Total    \\\midrule
Access Control      & \numprint{0} \numprint{0}\%    & \numprint{44} \numprint{0}\%   & \numprint{47} \numprint{0}\%   & \numprint{1076} \numprint{2}\% & \numprint{0} \numprint{0}\%    & \numprint{2} \numprint{0}\%    & \numprint{614} \numprint{1}\%  & \numprint{2356} \numprint{4}\% & \numprint{384} \numprint{0}\%  & \numprint{3801} \numprint{8}\% \\
Arithmetic          & \numprint{1} \numprint{0}\%    & \numprint{0} \numprint{0}\%    & \numprint{102} \numprint{0}\%  & \numprint{18515} \numprint{39}\% & \numprint{13922} \numprint{29}\% & \numprint{34306} \numprint{72}\% & \numprint{0} \numprint{0}\%    & \numprint{0} \numprint{0}\%    & \numprint{7430} \numprint{15}\% & \numprint{37597} \numprint{79}\% \\
Denial of Service      & \numprint{0} \numprint{0}\%    & \numprint{0} \numprint{0}\%    & \numprint{0} \numprint{0}\%    & \numprint{0} \numprint{0}\%    & \numprint{485} \numprint{1}\%  & \numprint{880} \numprint{1}\%  & \numprint{0} \numprint{0}\%    & \numprint{2555} \numprint{5}\% & \numprint{11621} \numprint{24}\% & \numprint{12419} \numprint{26}\% \\
Front Running       & \numprint{0} \numprint{0}\%    & \numprint{0} \numprint{0}\%    & \numprint{0} \numprint{0}\%    & \numprint{2015} \numprint{4}\% & \numprint{0} \numprint{0}\%    & \numprint{0} \numprint{0}\%    & \numprint{7217} \numprint{15}\% & \numprint{0} \numprint{0}\%    & \numprint{0} \numprint{0}\%    & \numprint{8161} \numprint{17}\% \\
Reentrancy          & \numprint{19} \numprint{0}\%   & \numprint{0} \numprint{0}\%    & \numprint{2} \numprint{0}\%    & \numprint{8454} \numprint{17}\% & \numprint{496} \numprint{1}\%  & \numprint{308} \numprint{0}\%  & \numprint{2033} \numprint{4}\% & \numprint{8764} \numprint{18}\% & \numprint{847} \numprint{1}\%  & \numprint{14747} \numprint{31}\% \\
Time Manipulation   & \numprint{0} \numprint{0}\%    & \numprint{0} \numprint{0}\%    & \numprint{90} \numprint{0}\%   & \numprint{0} \numprint{0}\%    & \numprint{1470} \numprint{3}\% & \numprint{1452} \numprint{3}\% & \numprint{0} \numprint{0}\%    & \numprint{1988} \numprint{4}\% & \numprint{68} \numprint{0}\%   & \numprint{4069} \numprint{8}\% \\
Unchecked Low Calls & \numprint{0} \numprint{0}\%    & \numprint{0} \numprint{0}\%    & \numprint{4} \numprint{0}\%    & \numprint{443} \numprint{0}\%  & \numprint{0} \numprint{0}\%    & \numprint{0} \numprint{0}\%    & \numprint{592} \numprint{1}\%  & \numprint{12199} \numprint{25}\% & \numprint{2867} \numprint{6}\% & \numprint{14656} \numprint{30}\% \\
Unknown Unknows    & \numprint{26} \numprint{0}\%   & \numprint{135} \numprint{0}\%  & \numprint{1032} \numprint{2}\% & \numprint{11126} \numprint{23}\% & \numprint{0} \numprint{0}\%    & \numprint{0} \numprint{0}\%    & \numprint{561} \numprint{1}\%  & \numprint{9133} \numprint{19}\% & \numprint{14113} \numprint{29}\% & \numprint{28355} \numprint{59}\% \\\midrule
Total               & \numprint{46} \numprint{0}\%   & \numprint{179} \numprint{0}\%  & \numprint{1203} \numprint{2}\% & \numprint{22994} \numprint{48}\% & \numprint{14665} \numprint{30}\% & \numprint{34764} \numprint{73}\% & \numprint{8781} \numprint{18}\% & \numprint{22269} \numprint{46}\% & \numprint{24906} \numprint{52}\% & \numprint{44589} \numprint{93}\% \\
\bottomrule

\end{tabularx}
\end{table*}

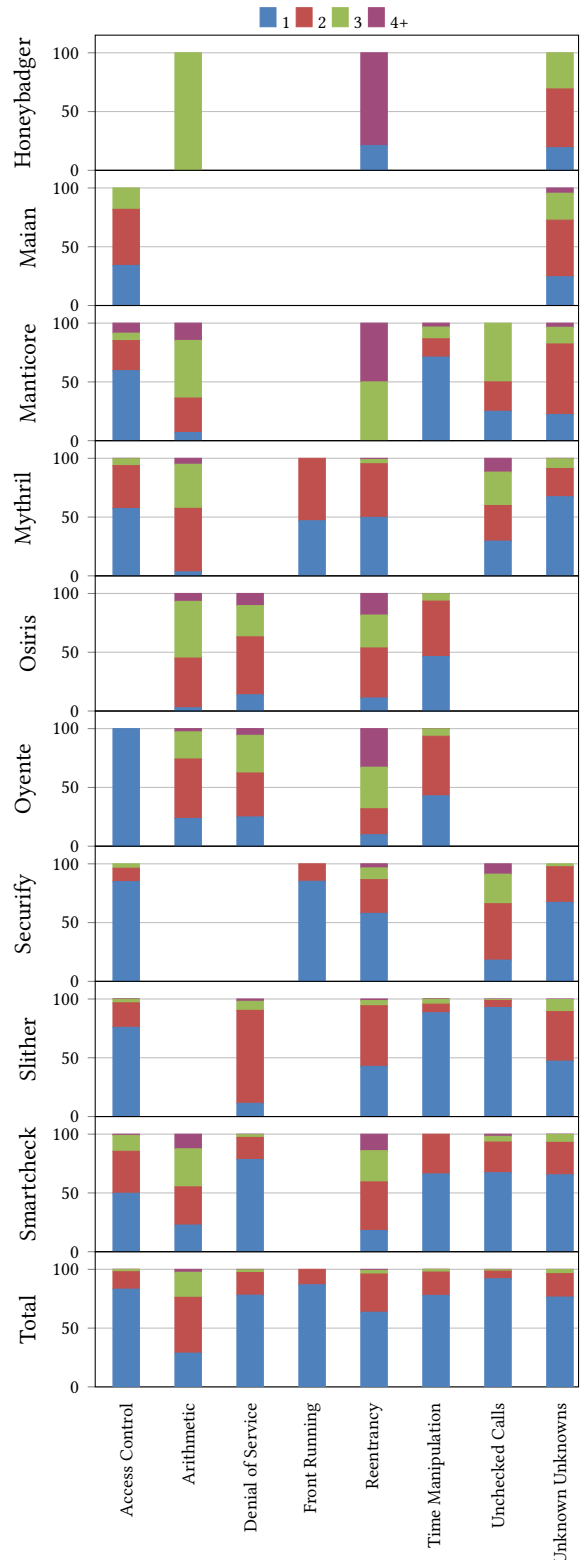
\begin{figure}
    \centering
    \begin{tikzpicture}
\begin{groupplot}[
    group style={
        group name=my plots,
        group size=1 by 10,
        xlabels at=edge bottom,
        xticklabels at=edge bottom,
        vertical sep=0pt
    },
    ybar stacked,
    footnotesize,
    width= 0.45*\textwidth,
    height=3.38cm,
    xmin=-0.5, xmax=7.3,
    ymin=0, ymax=115,
    % y filter/.expression={y<1.0 ? nan : y},
    major x tick style = transparent,
    ymajorgrids = true,
    scaled ticks=false,
    xtick={0,1,...,7},
    xticklabels={Access Control,Arithmetic,Denial of Service,Front Running,Reentrancy,Time Manipulation,Unchecked Calls,Unknown Unknowns},
    x tick label style={rotate=90,anchor=east},
    ylabel near ticks, 
    tickpos=left,
    ytick align=outside,
    xtick align=outside,
    legend style={
        at={(0.5,1.25)},anchor=north,
        legend columns=-1,
        draw=none,
        fill=none
    }
]

\nextgroupplot[ylabel = Honeybadger]

\addplot [style={bblue,fill=bblue,mark=none}] coordinates{(0,0.00)(1,0.00)(2,0.00)(3,0.00)(4,21.05)(5,0.00)(6,0.00)(7,19.23)};
\addplot [style={rred,fill=rred,mark=none}] coordinates{(0,0.00)(1,0.00)(2,0.00)(3,0.00)(4,0.00)(5,0.00)(6,0.00)(7,50.00)};
\addplot [style={ggreen,fill=ggreen,mark=none}] coordinates{(0,0.00)(1,100.00)(2,0.00)(3,0.00)(4,0.00)(5,0.00)(6,0.00)(7,30.77)};
\addplot [style={ppurple,fill=ppurple,mark=none}] coordinates{(0,0.00)(1,0.00)(2,0.00)(3,0.00)(4,78.95)(5,0.00)(6,0.00)(7,0.00)};

\legend{\strut 1, \strut 2, \strut 3, \strut 4+}

\nextgroupplot[ylabel = Maian]

\addplot [style={bblue,fill=bblue,mark=none}] coordinates{(0,34.09)(1,0.00)(2,0.00)(3,0.00)(4,0.00)(5,0.00)(6,0.00)(7,24.44)};
\addplot [style={rred,fill=rred,mark=none}] coordinates{(0,47.73)(1,0.00)(2,0.00)(3,0.00)(4,0.00)(5,0.00)(6,0.00)(7,48.15)};
\addplot [style={ggreen,fill=ggreen,mark=none}] coordinates{(0,18.18)(1,0.00)(2,0.00)(3,0.00)(4,0.00)(5,0.00)(6,0.00)(7,22.96)};
\addplot [style={ppurple,fill=ppurple,mark=none}] coordinates{(0,0.00)(1,0.00)(2,0.00)(3,0.00)(4,0.00)(5,0.00)(6,0.00)(7,4.44)};

\nextgroupplot[ylabel = Manticore]

\addplot [style={bblue,fill=bblue,mark=none}] coordinates{(0,59.57)(1,6.86)(2,0.00)(3,0.00)(4,0.00)(5,71.11)(6,25.00)(7,22.29)};
\addplot [style={rred,fill=rred,mark=none}] coordinates{(0,25.53)(1,29.41)(2,0.00)(3,0.00)(4,0.00)(5,15.56)(6,25.00)(7,59.98)};
\addplot [style={ggreen,fill=ggreen,mark=none}] coordinates{(0,6.38)(1,49.02)(2,0.00)(3,0.00)(4,50.00)(5,10.00)(6,50.00)(7,14.24)};
\addplot [style={ppurple,fill=ppurple,mark=none}] coordinates{(0,8.51)(1,14.71)(2,0.00)(3,0.00)(4,50.00)(5,3.33)(6,0.00)(7,3.49)};

\nextgroupplot[ylabel = Mythril]

\addplot [style={bblue,fill=bblue,mark=none}] coordinates{(0,57.25)(1,3.41)(2,0.00)(3,46.85)(4,49.76)(5,0.00)(6,29.57)(7,67.31)};
\addplot [style={rred,fill=rred,mark=none}] coordinates{(0,36.52)(1,54.03)(2,0.00)(3,53.15)(4,45.59)(5,0.00)(6,30.25)(7,24.01)};
\addplot [style={ggreen,fill=ggreen,mark=none}] coordinates{(0,5.86)(1,37.51)(2,0.00)(3,0.00)(4,3.56)(5,0.00)(6,28.44)(7,8.31)};
\addplot [style={ppurple,fill=ppurple,mark=none}] coordinates{(0,0.37)(1,5.06)(2,0.00)(3,0.00)(4,1.09)(5,0.00)(6,11.74)(7,0.37)};

\nextgroupplot[ylabel = Osiris]

\addplot [style={bblue,fill=bblue,mark=none}] coordinates{(0,0.00)(1,2.72)(2,13.81)(3,0.00)(4,11.09)(5,46.39)(6,0.00)(7,0.00)};
\addplot [style={rred,fill=rred,mark=none}] coordinates{(0,0.00)(1,42.34)(2,49.28)(3,0.00)(4,42.54)(5,47.14)(6,0.00)(7,0.00)};
\addplot [style={ggreen,fill=ggreen,mark=none}] coordinates{(0,0.00)(1,48.22)(2,26.60)(3,0.00)(4,28.02)(5,6.26)(6,0.00)(7,0.00)};
\addplot [style={ppurple,fill=ppurple,mark=none}] coordinates{(0,0.00)(1,6.72)(2,10.31)(3,0.00)(4,18.35)(5,0.20)(6,0.00)(7,0.00)};

\nextgroupplot[ylabel = Oyente]

\addplot [style={bblue,fill=bblue,mark=none}] coordinates{(0,100.00)(1,23.62)(2,24.89)(3,0.00)(4,9.74)(5,42.84)(6,0.00)(7,0.00)};
\addplot [style={rred,fill=rred,mark=none}] coordinates{(0,0.00)(1,50.55)(2,37.39)(3,0.00)(4,22.08)(5,50.69)(6,0.00)(7,0.00)};
\addplot [style={ggreen,fill=ggreen,mark=none}] coordinates{(0,0.00)(1,23.10)(2,32.05)(3,0.00)(4,35.39)(5,6.27)(6,0.00)(7,0.00)};
\addplot [style={ppurple,fill=ppurple,mark=none}] coordinates{(0,0.00)(1,2.73)(2,5.68)(3,0.00)(4,32.79)(5,0.21)(6,0.00)(7,0.00)};

\nextgroupplot[ylabel = Securify]

\addplot [style={bblue,fill=bblue,mark=none}] coordinates{(0,84.85)(1,0.00)(2,0.00)(3,85.16)(4,57.85)(5,0.00)(6,18.07)(7,67.20)};
\addplot [style={rred,fill=rred,mark=none}] coordinates{(0,11.40)(1,0.00)(2,0.00)(3,14.84)(4,28.68)(5,0.00)(6,47.97)(7,30.30)};
\addplot [style={ggreen,fill=ggreen,mark=none}] coordinates{(0,3.75)(1,0.00)(2,0.00)(3,0.00)(4,9.99)(5,0.00)(6,25.17)(7,2.50)};
\addplot [style={ppurple,fill=ppurple,mark=none}] coordinates{(0,0.00)(1,0.00)(2,0.00)(3,0.00)(4,3.49)(5,0.00)(6,8.78)(7,0.00)};

\nextgroupplot[ylabel = Slither]

\addplot [style={bblue,fill=bblue,mark=none}] coordinates{(0,75.89)(1,0.00)(2,11.23)(3,0.00)(4,42.70)(5,88.28)(6,92.77)(7,47.15)};
\addplot [style={rred,fill=rred,mark=none}] coordinates{(0,20.84)(1,0.00)(2,79.10)(3,0.00)(4,51.55)(5,7.34)(6,6.02)(7,42.09)};
\addplot [style={ggreen,fill=ggreen,mark=none}] coordinates{(0,3.10)(1,0.00)(2,7.71)(3,0.00)(4,4.42)(5,4.23)(6,0.79)(7,10.31)};
\addplot [style={ppurple,fill=ppurple,mark=none}] coordinates{(0,0.17)(1,0.00)(2,1.96)(3,0.00)(4,1.34)(5,0.15)(6,0.43)(7,0.45)};

\nextgroupplot[ylabel = Smartcheck]

\addplot [style={bblue,fill=bblue,mark=none}] coordinates{(0,49.74)(1,22.69)(2,78.45)(3,0.00)(4,18.06)(5,66.18)(6,67.18)(7,65.41)};
\addplot [style={rred,fill=rred,mark=none}] coordinates{(0,35.68)(1,32.49)(2,18.78)(3,0.00)(4,41.32)(5,33.82)(6,26.12)(7,27.61)};
\addplot [style={ggreen,fill=ggreen,mark=none}] coordinates{(0,13.54)(1,32.34)(2,2.33)(3,0.00)(4,26.56)(5,0.00)(6,4.88)(7,6.70)};
\addplot [style={ppurple,fill=ppurple,mark=none}] coordinates{(0,1.04)(1,12.48)(2,0.43)(3,0.00)(4,14.05)(5,0.00)(6,1.81)(7,0.29)};

\nextgroupplot[ylabel = Total]

\addplot [style={bblue,fill=bblue,mark=none}] coordinates{(0,83.16)(1,28.74)(2,78.03)(3,86.88)(4,63.52)(5,77.86)(6,91.99)(7,76.43)};
\addplot [style={rred,fill=rred,mark=none}] coordinates{(0,14.79)(1,47.45)(2,19.21)(3,13.12)(4,32.49)(5,19.81)(6,6.49)(7,19.89)};
\addplot [style={ggreen,fill=ggreen,mark=none}] coordinates{(0,1.95)(1,21.31)(2,2.36)(3,0.00)(4,3.09)(5,2.26)(6,1.17)(7,3.54)};
\addplot [style={ppurple,fill=ppurple,mark=none}] coordinates{(0,0.11)(1,2.49)(2,0.40)(3,0.00)(4,0.90)(5,0.07)(6,0.35)(7,0.14)};

\end{groupplot}
\end{tikzpicture}
\vspace{-0.3em}
    \caption{Proportion of vulnerabilities identified by exactly one (1), two (2) or three (3) tools, and by four tools or more (4+).}
    \label{fig:smartbugs:proportion_more_than_one_tool}
\end{figure}

To answer the second research question, we analyzed the ability of the \nbTools selected tools to detect vulnerabilities in the contracts from the dataset \sbwild (described in \autoref{sec:smartbugs:real_contracts}).
We followed the same methodology as in the previous research question, however, for \sbwild, we do not have an oracle to identify the vulnerabilities. 

\autoref{tab:smartbugs:wild-all-detection} presents the results of executing the \nbTools tools on the \nbContracts contracts. It shows that the \nbTools tools are able to detect eight different categories of vulnerabilities.
Note that the vulnerabilities detected by \textit{HoneyBadger} are contracts that look vulnerable but are not. They are designed to look vulnerable in order to steal Ether from people that tries to exploit the vulnerability.
In total, \numprint{44589} contracts (93\%) have at least one vulnerability detected by one of the \nbTools tools.

Such a high number of vulnerable contracts suggests the presence of a considerable number of false positives.
\textit{Oyente} is the approach that identifies the highest number of contracts as vulnerable (73\%), mostly due to vulnerabilities in the \textit{Arithmetic} category.
This observation is coherent with the observation of Parizi \textit{et~al.}~\cite{Parizi:2018:EVA:3291291.3291303}, since they determine that \textit{Oyente} has the highest number of false positives when compared to \textit{Mythril}, \textit{Securify}, and \textit{Smartcheck}.

Since we observed a potentially large number of false positives, we analyzed to what
extent the tools agree in vulnerabilities they flag. The hypothesis is that if a vulnerability is identified exclusively by a single tool, the probability of it 
being a false positive increases.
\autoref{fig:smartbugs:proportion_more_than_one_tool} presents the results of this analysis. This figure shows the proportion of detected vulnerabilities that have been
identified exclusively by one tool alone, two tools, three tools, and finally by four 
or more tools.
\textit{HoneyBadger} has a peculiar, but useful role: if \textit{HoneyBadger} detects a vulnerability, it actually means that the vulnerability does not exist. So, consensus with \textit{HoneyBadger} suggests the presence of false positives.

It is clear from the figure that the large majority of the vulnerabilities have been
detected by one tool only. One can observe that there are 71.25\% of the
\textit{Arithmetic} vulnerabilities found by more than one tool.
It is also the category with the highest consensus between four and more tools: 937
contracts are flagged as having an \textit{Arithmetic} vulnerability with a consensus 
of more than three tools.
It is followed by the \textit{Reentrancy} category with 133 contracts receiving a
consensus of four tools or more. These results suggest that combining several of these
tools may yield more accurate results, with fewer false positives and negatives. 

The tool \textit{HoneyBadger} is different: instead of detecting vulnerabilities, it detects malicious contracts that try to imitate vulnerable contracts in order to attract transactions to their honeypots. %malicious contract.
Therefore, when \textit{HoneyBadger} is detecting a \textit{Reentrancy} vulnerability, it means that the contract looks vulnerable to \textit{Reentrancy} but it is not.
\autoref{fig:smartbugs:proportion_more_than_one_tool} shows that 15 contracts identified by \textit{HoneyBadger} with vulnerabilities of type \textit{Reentrancy} have been detected by three other tools as \textit{Reentrancy} vulnerable.

\begin{figure}
    \centering

    \begin{adjustbox}{clip,trim=0 0cm 0cm 3em}
    \scalebox{0.6}{
        \input{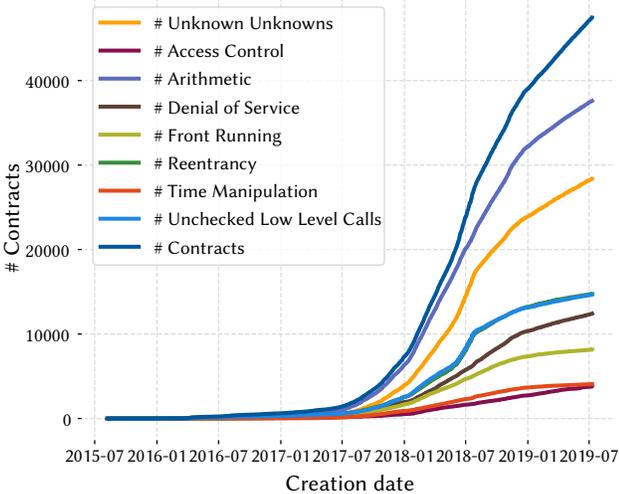}
    }
    \end{adjustbox}
    \caption{Evolution of number of vulnerabilities over time.}
    \label{fig:smartbugs:category_evolution}
\end{figure}

We also analyzed the evolution of the vulnerabilities over time.
\autoref{fig:smartbugs:category_evolution} presents the evolution of the number of vulnerabilities by category.
It firstly shows that the total number of unique contracts started to increase exponentially at the end of 2017 when Ether was at its highest value.
Secondly, we can observe two main groups of curves. The first one contains the categories \textit{Arithmetic} and \textit{Unknown Unknowns}. These two categories follow the curve of the total number of contracts.
The second group contains the other categories. The growing number of vulnerable contracts seems to slow down from July 2018.
Finally, this figure shows that the evolution of categories \textit{Reentrancy} and \textit{Unchecked Low Level Calls} is extremely similar (the green line of \textit{Reentrancy} is also hidden by the blue line of \textit{Unchecked Low Level Calls}).
This suggests a correlation between vulnerabilities in these two categories. % reentrancy and unchecked low-level calls vulnerabilities.

\begin{figure}
    \centering
    \includegraphics[width=.475\textwidth]{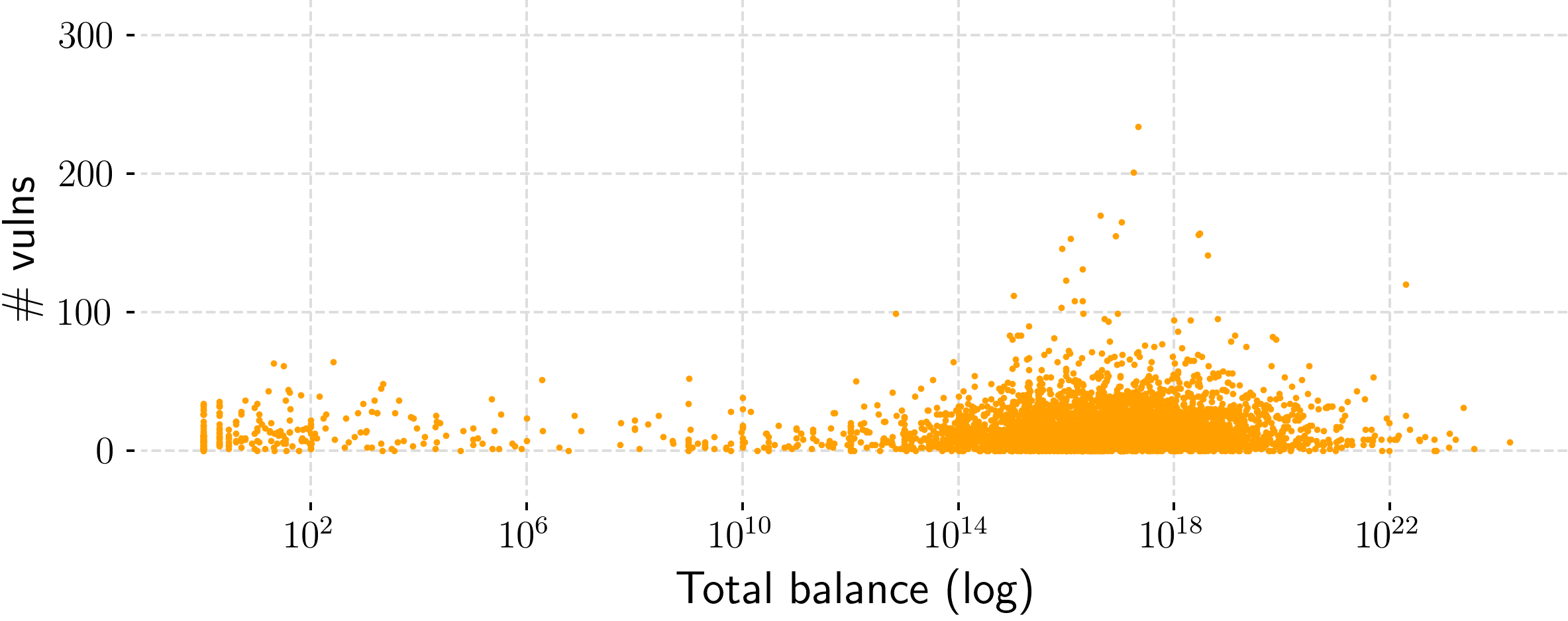}
    \caption{Correlation between the number of vulnerabilities and balance in Wei (one Ether is $10^{18}$ Wei).}
    \label{fig:smartbugs:vuln_balance}
\end{figure}

And lastly, \autoref{fig:smartbugs:vuln_balance} presents the correlation between the number 
of vulnerabilities and the balance of the contracts.  It shows that the contracts that have 
a balance between $10^{14}$ Wei and $10^{20}$ Wei have more vulnerabilities than other
contracts. Hence, the richest and the \textit{middle class} seem to be less impacted. Per
category, we have not observed any significant differences worth reporting.

\answer{2}{
\textbf{How many vulnerabilities are present in the Ethereum blockchain?}
The \nbTools tools identify vulnerabilities in 93\% of the contracts, which suggests a high number of false positives. \textit{Oyente}, alone, detects vulnerabilities in 73\% of the contracts.
By combining the tools to create a consensus, we observe that only a few number of vulnerabilities received a consensus of four or more tools: 937 for \textit{Arithmetic} and 133 for \textit{Reentrancy}.
}

\subsection{Execution Time of the Analysis Tools (RQ3)}

In this section, we present the execution time required by the tools to analyze the \nbContracts of the \sbwild dataset (see \autoref{sec:smartbugs:real_contracts}).
In order to measure the time of the execution, we recorded for each individual analysis when it started and when it ended. 
The duration of the analysis is the difference between the starting time and the ending time. 
An individual execution is composed of the following steps:
1) start the Docker image and bind the contract to the Docker instance;
2) clean the Docker container; and
3) parse and output the logs to the results folder.

\autoref{tab:smartbugs:execution_time} presents the average and total times used by each tool. 
%\red{As presented in \autoref{sec:smartbugs:execution}, we used two cloud providers: Scaleway and Google Cloud. With Scaleway, we used three servers with 32 vCPUs and 128 GB of RAM, and with Google Cloud, we used three servers with 32 vCPUs with 30GB of RAM. In total, the execution of all the tools required \ExecDuration.} 
The average execution time is per contract. It considers the execution of the tool on a contract, including compilation, construction of IR/graphs, analysis and parsing the results.
In the table, we can observe three different groups of execution time: the tools that take a few seconds to execute, the tools that take a few minutes, and \textit{Manticore} that takes 24 minutes.
\textit{Oyente}, \textit{Osiris}, \textit{Slither}, \textit{Smartcheck}, and \textit{Solhint} are much faster tools that take between 5 and 30 seconds on average to analyze a smart contract.
\textit{HoneyBadger}, \textit{Maian}, \textit{Mythril}, and \textit{Securify} are slower and take between 1m24s and 6m37s to execute. Finally, Manticore takes 24m28s.
The difference in execution time between the tools is dependent on the technique that each tool uses. 
Pure static analysis tools such as \textit{Smartcheck} and \textit{Slither} are fast since they do not need to compile nor execute contracts 
%only analyze the AST of the contract 
to identify vulnerabilities and bad practices.

\textit{Securify}, \textit{Maian}, \textit{Mythril}, and \textit{Manticore} analyze the EVM 
bytecode of the contracts. It means that those tools require the contract to be compiled 
before doing the analysis.
The additional compilation step slows down the analysis.
\textit{Manticore} is the slowest of all the tools because this tool only analyzes an 
internal contract at a time (Solidity source files can contain an arbitrary number of 
contract definitions).
Consequently, this tool has the major drawback of having the compilation overhead for each 
internal contract that it analyzes.

The average execution time does not reflect the complete picture of the performance of a tool. For example, \textit{Maian} and \textit{Manticore} use several processes, and \textit{Maian} uses up to 16GB of RAM. Consequently, \textit{Maian} and \textit{Manticore} are difficult to parallelize.  
We were able to run only four, and ten parallel executions for respectively \textit{Maian} and \textit{Manticore} on a 32-core server with 30GB of RAM.
This also explains why we were not able to execute those two tools on the complete dataset of \nbContracts smart contracts.

Interestingly, the slowest tools do not have better accuracy (see \autoref{sec:smartbugs:rq1}). \textit{Mythril}, for example, which has the best accuracy according to our evaluation, takes on average of 1m24s to analyze a contract.
It is much faster than \textit{Manticore} that only has an accuracy of 11\% compared to the 27\% of \textit{Mythril}.

The execution time of \textit{Maian} is surprising compared to the results that have been presented in the \textit{Maian} paper \cite{nikolic2018finding}. 
Indeed, the authors claimed that it takes on average 10 seconds to analyze a contract while we observe that it takes 5m16s in our experiment on similar hardware.
The difference in execution times can potentially be explained by the difference of input uses in the two experiments. We use the source code of the contract as input, and \textit{Maian}'s authors use the bytecode. 
The overhead for the compilation of the contract seems to be the major cost of execution for this tool.

\begin{table}[t]
    \small
    \caption{Average execution time for each tool.}
    \label{tab:smartbugs:execution_time}
    \centering
    \begin{tabularx}{0.47\textwidth}{@{}r X l r @{}}\toprule
\multirow{2}{*}{\#} & \multirow{2}{*}{Tools} & \multicolumn{2}{c}{Execution time}  \\
    &            & Average & Total  \\\midrule
1 & Honeybadger & 0:01:38    & 23 days, 13:40:00 \inlinechart{1987200}{48729600}\\
2 & Maian      & 0:05:16    & 49 days, 10:06:15 \inlinechart{4233600}{48729600}\\
3 & Manticore  & 0:24:28    & 184 days, 01:59:02 \inlinechart{15897600}{48729600}\\
4 & Mythril    & 0:01:24    & 46 days, 07:46:55 \inlinechart{3974400}{48729600}\\
5 & Osiris     & 0:00:34    & 18 days, 10:19:01 \inlinechart{1555200}{48729600}\\
6 & Oyente     & 0:00:30    & 16 days, 04:50:11 \inlinechart{1382400}{48729600}\\
7 & Securify   & 0:06:37    & 217 days, 22:46:26 \inlinechart{18748800}{48729600}\\
8 & Slither    & 0:00:05    & 2 days, 15:09:36 \inlinechart{172800}{48729600}\\
9 & Smartcheck & 0:00:10    & 5 days, 12:33:14 \inlinechart{432000}{48729600}\\
 \midrule
    \multicolumn{2}{l}{\textbf{Total}} & 0:04:31    & 564 days, 3:10:39 \\
\bottomrule
    \end{tabularx}
\end{table}

\answer{3}{\textbf{How long do the tools require to analyze the smart contracts?}
On average, the tools take 4m31s to analyze one contract. 
However, the execution time largely varies between the tools.
\textit{Slither} is the fastest tool and takes on average only 5 seconds to analyze a contract.
\textit{Manticore} is the slowest tool. It takes on average 24m28s to 
analyze a contract. We also observe that the execution speed is not the only factor 
that impacts the performance of the tools. \textit{Securify} took more time to execute than
\textit{Maian}, but \textit{Securify} can easily be parallelized and therefore analyze the 
\nbContracts\ contracts much faster than \textit{Maian}. Finally, we have not observed a correlation 
between accuracy and execution time.
}

\section{Discussion}
\label{sec:smartbugs:discussion}
We discuss the practical implications of our findings from
the previous section, as well as outline the potential threats to
their validity.
\subsection{Practical Implications and Challenges}
\label{sec:smartbugs:challenges}
Despite the advances in automatic analysis tools of smart contracts during the
last couple of years, the practical implication our study highlights is that
there remain several open challenges to be tackled by future
work. We identify four core challenges: increasing and ensuring
the quality of the analysis, extending the scope of problems addressed
by these tools, integrating the analysis into the development process, and 
extending the current taxonomy.

\paragraph{Quality:} This challenge is about increasing the likelihood that a tool 
identifies real vulnerabilities, yielding close to zero false positives and false 
negatives. Our study demonstrates that this is far from being the case, and 
future work should be invested in improving the quality of the tools. Addressing 
this challenge is perhaps an important step toward real-life adoption of these 
tools and techniques.

\paragraph{Scope:} Although there might not be a technique that finds all sorts 
of vulnerabilities, this challenge is about further extending existing techniques 
so that more real vulnerabilities can be found. In the previous section, we briefly
discussed  a potential way to address this challenge: crafting a novel technique
combining  complementary tools. Combining static with dynamic analysis might also 
be an interesting avenue for future work. 

\paragraph{Development process:} This challenge is about integrating these tools 
into the development process, thus contributing to real-life adoption. 
To ease the interaction of developers with these tools, hence making them
useful during the development life-cycle, the following could bring added value: 
integration with other orthogonal techniques (such as bug detection tools, dynamic
analysis techniques, and generic linters), integration with popular IDEs, 
interactive reports (e.g., highlight vulnerable code), and explainable warnings. 

\paragraph{Taxonomy:} Another practical implication of our work is that the 
current state-of-the-art set of 10 categories in DASP10 does not seem to be 
comprehensive enough to cover all vulnerabilities that affect smart contracts 
deployed in Ethereum. DASP10 includes the category \textit{Unknown Unknowns}, because as 
the creators of the DASP taxonomy observed, \textit{as long as investors decide 
to place large amounts of money on complex but lightly-audited code, we will 
continue to see new discoveries 
leading to dire consequences}. Our work sheds light on potential new categories that
could extend DASP10, such as \textit{Dependence on environment data} and 
\textit{Locked Ether}.
The latter could include not only the cases where Ether is locked indefinitely, but 
also cases of \textit{Honeypots} as defined by Torres \textit{et~al.}~\cite{honeybadger} (since Ether becomes locked, except for the attacker).
Our findings suggest new categories comparable to those proposed in the recent survey by Chen~\textit{et~al.}~\cite{chen2019survey} (available online on August 13, 2019).

\subsection{Threats to Validity}
\label{sec:smartbugs:threats_to_Validity}
A potential threat to the internal validity is that, due to the complexity of the
\smartbugs{} framework, there may remain an implementation bug somewhere in the 
codebase. We extensively tested the framework to mitigate this risk. Furthermore, 
the framework and the raw data are publicly available for other researchers
and potential users to check the validity of the results.

A potential threat to the external validity is related to the fact that the set of 
smart contracts we have considered in this study may not be an accurate 
representation of the set of vulnerabilities that can happen during development. 
We attempt to reduce the selection bias by leveraging a large collection of real,
reproducible smart contracts. Another potential threat is that we may have missed 
a tool or failed to reproduce a tool that excels all other tools. To mitigate this 
risk, we contacted the authors of the tools if no source code was found. We also 
aim to reduce threats to external validity and
ensure the reproducibility of our evaluation by providing the source of our 
instrumentation tool, the scripts used to run the evaluation, and all data gathered.

A potential threat to the construct validity relates to the fact that we needed to 
manually label the vulnerable smart contracts into one of the DASP categories,
and these could be mislabeled. This risk was mitigated as follows: all authors labeled
the contracts, and then disagreements were discussed to reach a consensus. Another 
potential threat to the validity is the timeout used for the analysis: 30 minutes. 
This threat was mitigated by executing all the tools on the \sbcurated{} dataset.

%\todo{his uncontrolled dataset is however a valuable experiment because it gives a sense of the potential noise that tools might generate when analyzing contracts. In such a big dataset, the discussion on the number of FPs isn’t trivial, and we decided to use consensus as a proxy. This is a threat, and we’ll make it clearer in S4.2.}
The analysis based on the dataset \sbwild\ is valuable for it gives a sense of the potential noise that tools might generate when analysing contracts. However, with such a large dataset, the discussion on the number of false positives is challenging and there is a risk that too many false positives are identified. To mitigate this risk, we decided to use consensus as a proxy: the hypothesis is that the greater the number of tools identifying a vulnerability, the more likely that vulnerability is a true positive.
\balance
\section{Related Work}
\label{sec:smartbugs:related-works}
As discussed in \autoref{sec:smartbugs:tools}, there are several automated analysis tools available. Notwithstanding, despite the recent increase 
interest in the analysis of smart contracts, to the best of our knowledge, our 
work is the first systematic comparison of recently proposed techniques to better 
understand their real capabilities. 

\paragraph{Datasets and Repositories:}
Reproducibility is enabled by a benchmark containing smart contracts that 
other researchers can use \textit{off-the-shelf}. 
There are just a few repositories of smart contracts available to the
research community, such as \emph{VeriSmartBench}\footnote{VeriSmartBench: \url{https://github.com/soohoio/VeriSmartBench}}, \emph{evm-analyzer-benchmark-suite}\footnote{evm-analyzer-benchmark-suite: \url{https://github.com/ConsenSys/evm-analyzer-benchmark-suite}}, \emph{EthBench}\footnote{EthBench: \url{https://github.com/seresistvanandras/EthBench}}, \emph{smart-contract-benchmark}\footnote{smart-contract-benchmark: \url{https://github.com/hrishioa/smart-contract-benchmark}}, and \emph{not-so-smart-contracts}\footnote{not-so-smart-contracts: \url{https://github.com/crytic/not-so-smart-contracts}}. 
These, however, are essentially collections of contracts and are not designed to enable
reproducibility nor to facilitate comparison of research. Our dataset \sbcurated offers a known
vulnerability taxonomy, positioning itself as a reference dataset to the research 
community. 

\paragraph{Empirical studies:}
Chen \textit{et~al.}~\cite{Jiachi19:1905.01467} discussed an empirical study on code
smells for Ethereum smart contracts. Based on posts from Stack Exchange and 
real-world smart
contracts, they defined 20 distinct code smells for smart contracts and categorized 
them into security, architecture, and usability issues. Furthermore, they manually 
labeled a dataset of smart contracts.\footnote{CodeSmell:
\url{https://github.com/CodeSmell2019/CodeSmell}}
Pinna \textit{et~al.}~\cite{pinna2019massive} performed a comprehensive empirical study of 
smart contracts deployed on the Ethereum blockchain with the objective to provide
an overview of smart contracts features, such as type of transactions, the role of 
the development community, and the source code characteristics.
Parizi \textit{et~al.}~\cite{Parizi:2018:EVA:3291291.3291303} carried out an experimental
assessment of static smart contracts security testing tools. They tested Mythril, Oyente, Securify, and Smartcheck on ten real-world smart contracts. Concerning the accuracy of the tools, Mythril was found to be the most accurate. Our results corroborate the findings, but in a more systematic and comprehensive manner.

\paragraph{Execution Frameworks:} Execution frameworks to simplify and automate the
execution of smart contract analysis tools are scarce. Solhydra \cite{Solhydra} is a 
CLI tool to analyze Solidity smart contracts with several static analysis tools. It
generates a report with results of the tool analysis. Unlike \smartbugs, which was
designed to ease the addition of new analysis tools, Solhydra does not offer this
flexibility. Furthermore, Solhydra has not been updated in more than a year.

\section{Conclusion}\label{sec:smartbugs:conclusions}

In this paper, we presented an empirical evaluation of \nbTools automated analysis tools on \nbCurated annotated vulnerable contracts and on \nbContracts contracts taken from the Ethereum’s network.
The goal of this experiment was to obtain an overview of the current state of automated analysis tools for Ethereum smart contracts.
During this empirical evaluation, we considered all available smart contracts analysis tools and all the available Ethereum contracts that have at least one transaction. We used the DASP10 category of smart contract vulnerabilities as the reference to classify vulnerabilities.

We found out that the current state of the art is not able to detect vulnerabilities from two categories of DASP10: \textit{Bad Randomness} and \textit{Short Addresses}. Also, the tools are only able to detect together 48/115 (42\%) of the vulnerabilities from our \sbcurated dataset. \textit{Mythril} is the tool that has the higher accuracy and is able to detect 31/115 (27\%) of the vulnerabilities.

During the evaluation of the \nbTools tools on \sbwild, we observe that 97\% of the contracts are identified as vulnerable.
This suggests a considerable number of false positives. \textit{Oyente} plays an important role in this, since it detects vulnerabilities in 73\% of the contracts, mostly due to \textit{Arithmetic} vulnerabilities (72\%).
Finally, we observe that the tools (4 or more) succeed to find a consensus for 937 \textit{Arithmetic} vulnerabilities and 133 \textit{Reentrancy} vulnerabilities.

We argue that the execution framework and the two new datasets presented here are valuable assets for driving reproducible research in automated analysis of smart contracts.

% No space for future work?
%and we plan to keep improving it. Some of the next steps include i) the conversion of the tools' output into a standard format so that we can automatically compare the results of different tools and produce comparison reports, and \todo{What else?}

\section*{Acknowledgments}

This work has been co-funded by the European Union's Horizon 2020 research and innovation programme under the QualiChain project, Grant Agreement No 822404 and 
supported by national funds through FCT, Fundação para a Ciência e a Tecnologia, under projects UIDB/50021/2020 and PTDC/CCI-COM/29300/2017.
% This work was supported by national funds through FCT, Fundação para a Ciência e a Tecnologia, under project UIDB/50021/2020 and
%by Fundação para a Ciência e a Tecnologia (FCT), with the reference PTDC/CCI-COM/29300/2017.
%The work reported in this article was supported by national funds through Fundação para a Ciência e a Tecnologia (FCT) with reference UID/CEC/50021/2019

\bibliographystyle{ACM-Reference-Format}

\bibliography{references}

\end{document}